\documentstyle[aps,titlepage,psfig]{revtex}
\title{Dielectric function of nonideal plasmas and electrical dc
  conductivity}

\author{G. R\"opke}

\address{FB Physik, Universit\"at Rostock, 18051 Rostock, Germany}

\date{\today}

\newcommand{\be}{\begin{equation}}
\newcommand{\ee}{\end{equation}}
\newcommand{\ba}{\begin{eqnarray}}
\newcommand{\ea}{\end{eqnarray}}

\def\e{{\rm e}}

\begin{document}

\maketitle

\vspace{2cm}

\begin{abstract}
Within generalized linear response theory, an expression for the
dielectric function is derived which is consistent with standard
approaches to the electrical dc conductivity. Explicit results
are given for the first moment Born approximation. Some exact
relations as well as the limiting behaviour at small values of
the wave number and frequency are investigated.
\vspace{3mm}
\\
\end{abstract}

\section{Introduction}

The dielectric function $\epsilon(\vec k, \omega)$ describing the
response of a charged particle system to an external, time and space
dependent electrical field is related to various phenomena such as
electrical conductivity and optical absorption of light.  In
particular, it is an important quantity for plasma diagnostics, see,
e.g., recent applications to determine the parameters of picosecond
laser produced high-density plasmas \cite{S}.  However, the
application of widely used simplified expressions for the dielectric
function is questionable in the case of nonideal plasmas.

As well known, the electrical dc conductivity of a charged particle
system should be obtained as a limiting case of the dielectric
function. However, at present both quantities are treated by different
theories. A standard approach to the dc electrical conductivity is
given by the Chapman-Enskog approach \cite{ChapCow}. In dense
plasmas, where many-particle effects are of importance, linear
response theory has been worked out to relate the conductivity to
equilibrium correlation functions which can be evaluated using the
method of thermodynamic Green functions, see \cite{R}. This way it is
possible to derive results for the conductivity of partially
ionized plasmas not only on the level
of ordinary kinetic theory, but to include two-particle
nonequilibrium correlations as well\cite{RR}.

On the other hand, the dielectric function can also be expressed
in terms of equilibrium correlation functions, but the systematic
perturbative treatment to include collision effects is difficult
near the point $\vec k=0,\,\,\omega=0$, because an essential
singularity arises in zeroth order. Different possibilities
are known to go beyond the well-known RPA result. In the static limit,
local field corrections have been discussed extensively \cite{I}, and
the dynamical behavior of the corrections to the RPA in the
long-wavelength limit was investigated in the time-dependent mean
field theory neglecting damping effects \cite{GK}, see also
\cite{KG} for the strong coupling case. At arbitrary $\vec k,\,\,
\omega$, approximations are made on the basis of sum rules for the
lowest moments \cite{A}. However, these approximations cannot
give an unambiguous expression for $\epsilon(\vec k, \omega)$ in
the entire $\vec k, \,\, \omega$ space.

We will give here a unified approach to the dielectric function
as well as the dc conductivity, which is consistent with the
Chapman-Enskog approach to the dc conductivity and
which allows for a perturbation expansion also in the region of
small $\vec k,\,\,\omega$. In the following Section II the method
of generalized linear response \cite{Z} is presented which allows
to find very general relations between a dissipative
quantity and correlation functions describing the dynamical behaviour
of fluctuations in equilibrium. A special expression for the dielectric 
function is given in Section III  which is related
to the use of the force-force correlation function in evaluating
the conductivity.

Different methods can be applied to evaluate equilibrium correlation
functions for nonideal plasmas. We will use perturbation theory to
evaluate thermodynamic Green functions \cite{KKER}. Results in Born
approximation are given in Section IV. Using diagram techniques,
partial summations can be performed as shown in Ref. \cite{R}. An
alternative to evaluate equilibrium correlation function in strongly
coupled plasmas is given by molecular dynamical simulations. It is
expected that reliable results for the dielectric function for dense
systems by quantum molecular dynamics will be available in the near
future. Works in this direction are in progress but will not be
discussed in this paper.

To illustrate the general approach, explicit results for the
dielectric function in lowest moment Born approximation are given
for a Hydrogen plasma in Section V.
A sum rule as well as the conductivity are discussed.
The simple approximation considered here will be improved in a
subsequent paper \cite{RW}, where a four-moment approach to the
two-component plasma is investigated.

\section{Dielectric function within generalized linear response
  theory}

We consider a charge-neutral plasma consisting of two components with
masses $m_c$ and charges $e_c$, where the index $c$ denotes species
(electron $e$, ion $i$) and spin, under the influence of an external
potential $U_{\rm ext}(\vec r, t) = \e^{i(\vec k \vec r-\omega t)}
U_{\rm ext}(\vec k, \omega)$ + c.c. The total Hamiltonian $H_{\rm
  tot}(t)= H + H_{\rm ext}(t)$ contains the system Hamiltonian
\be
\label{1}
H=\sum_{c,p}E^c_p\,\, c^+_p c^{}_p + {1 \over 2} \sum_{cd,pp'q} V_{cd}(q)
\,\,c^+_{p-q}\, d^+_{p'+q}\,d^{}_{p'}\,c^{}_p
\ee
and the interaction with the external potential
\be
\label{2}
H_{\rm ext}(t) =U_{\rm ext}(\vec k, \omega)\, \e^{-i \omega t}\,
\sum_{c,p} e_c n^c_{p,-k} +\, {\rm c.c.},
\ee
where $E^c_p=\hbar^2p^2 /2m_c$ denotes the kinetic energy, $V_{cd}(q)
= e_ce_d/(\epsilon_0 \Omega_0 q^2)$ the Coulomb interaction and
$\Omega_0$ the normalization volume. Furthermore we introduced the
Wigner transform of the single-particle density
\be
\label{3}
n^c_{p,k}= \left(n^c_{p,-k}\right)^+=c^+_{p-k/2}\,\,c^{}_{p+k/2}.
\ee

Under the influence of the external potential, a time-dependent charge
density
\be
\label{4}
{1 \over \Omega_0} \sum_{c,p,k'}e_c \, \langle
\delta n^c_{p,k'} \rangle^t\,\, {\rm e}^{i \vec k' \vec r} +{\rm c.c.} =
{1 \over \Omega_0} \sum_{c,p}e_c\,\, \delta f_c(\vec p;\vec k,\omega)
\,\,{\rm e}^{i (\vec k \vec r-\omega t)} +{\rm c.c.}
\ee will be induced. Here, $\delta n^c_{p,k'}= n^c_{p,k'} - {\rm Tr}
\left\{n^c_{p,k'}\,\, \rho_0  \right\}$ denotes the deviation from
equilibrium given by
\be
\label{5}
\rho_0 = \exp (-\beta H + \beta \sum_c\mu_cN_c)\,\Big/\,{\rm Tr}
\exp (-\beta H + \beta \sum_c\mu_c N_c)\,\,.
\ee
The average $\langle \dots \rangle^t={\rm Tr} \left\{\dots \rho(t)
\right\} $ has to be performed with the nonequilibrium statistical
operator $\rho(t)$, which is derived in linear response
with respect to the external potential in Appendix A. For homogeneous
and isotropic systems, we find simple algebraic relations between the
different modes $(\vec k, \omega)$ of the external potential $U_{\rm
  ext}(\vec k, \omega)$ and the induced single-particle distribution
\be
\label{6}
\delta f_c(\vec p; \vec k,\omega) =
 \e^{i \omega t} \,\,\langle \delta n^c_{p,k} \rangle^t
\ee
which allow to introduce the dielectric
function $\epsilon (k, \omega)$, the electrical conductivity
$\sigma (k, \omega)$, and the polarization function $\Pi (k,
\omega)$. From standard electrodynamics we have
\ba
\label{7}
\epsilon ( k, \omega)& =& 1+ {i \over \epsilon_0 \omega}\,\, \sigma
(k, \omega) = 1 - {1 \over \epsilon_0 k^2}\,\, \Pi (k,
\omega)\,\,, \nonumber \\
\Pi (k, \omega)&=& {1 \over \Omega_0} \sum_{c,p} e_c \,\,\delta
f_c(\vec p; \vec k, \omega)\,\, {1 \over U_{\rm eff}(k,\omega)}
\ea
with $U_{\rm eff}(k, \omega)=U_{\rm ext}(k, \omega)/\epsilon(k,
\omega)$.
Using the equation of continuity
\be
\label{8}
\omega \sum_p \delta f_c(\vec p; \vec k,\omega)= {k \over m_c} \sum_p \hbar
p_z\,\,\delta f_c(\vec p; \vec k,\omega)\,\,,
\ee
where the $z$ direction is parallel to $\vec k$,
$\vec k = k \vec e_z$, we can also express
\ba
\label{9}
\Pi (k, \omega)&=& {k \over \omega}\,\,{1 \over \Omega_0} \sum_{c,p} {e_c
  \over m_c}\,\, \hbar p_z \,\,\delta
f_c(\vec p; \vec k, \omega)\,\, {1 \over U_{\rm eff}(k,\omega)}\nonumber \\
&=& {k \over \omega} \,\,\langle J_k \rangle^t\,\, \e^{i \omega t}\,\,
{1 \over U_{\rm eff}(k,\omega)}
\ea
with the current density operator
\be
\label{10}
J_k={1 \over \Omega_0}\,\, \sum_{c,p} {e_c
  \over m_c}\,\, \hbar p_z\,\, n_{p,k}^c\,\,.
\ee

The main problem in evaluating the mean value of the current density $
\langle J_k \rangle^t  $, Eq. (\ref{10}), is the determination of
$\rho(t)$. In linear response theory
where the external potential is considered to be weak, the statistical
operator $\rho(t)$ can be found as shown in Appendix A. An important
ingredient is that a set of relevant observables $\{ B_n \}$ can be
introduced whose mean values $\langle B_n \rangle^t$ characterize the
nonequilibrium state of the
system. The non-equilibrium statistical operator is constructed using
a corresponding set of thermodynamic parameters $\phi_n(t)$. For weak
perturbations, in linear response theory it is assumed that the
$\phi_n(t)$ are linear with respect to the external potential, and a set
of generalized response equations is derived which allow to evaluate
the response parameters $\phi_n(t)$. The coefficients of
these response equations are given in terms of equilibrium correlation
functions which can be evaluated using the methods of quantum statistics.

Solving this set of linear response equations by using Cramers rule,
the response parameters can be eliminated. If the current density
operator $J_k$ can be represented as a superposition of the relevant
observables $B_n$, we find
\be
\label{11}
\Pi(k, \omega)= i\, {k^2 \over \omega}\,\,\beta\, \Omega_0
\left| \begin{array}{rr}   0 & M_{0n} (k,\omega)  \\
  M_{m0}(k,\omega)  & M_{mn}(k,\omega)
\end{array} \right| \Big/\, |M_{mn} (k,\omega) |
\ee
with
\ba
\label{12}
M_{0n}(k,\omega)& =& (J_{k};B_{n})\,\,,\qquad
M_{m0}(k,\omega) = (B_{m}; \hat{J}_{k})\,\,, \nonumber \\
M_{mn}(k,\omega)& =& (B_{m}; [\dot B_{n}-i \omega B_{n}]) +
\langle \dot B_{m};
[\dot B_{n}-i \omega B_{n}]\rangle_{\omega+i \eta}
- \frac{\langle \dot B_{m};
{J}_{k}\rangle_{\omega+i \eta}}{\langle B_{m};
{J}_{k}\rangle_{\omega+i \eta}}\,\, \langle B_{m};[\dot B_{n}-
i \omega B_{n}] \rangle_{\omega+i \eta}\,\,.
\ea

The equilibrium correlation functions are defined as
\ba
\label{13}
(A;B)&=& (B^+;A^+)={1 \over \beta} \int_0^\beta d\tau\,\, {\rm Tr} \left[
A(-i \hbar \tau) B^+ \rho_0 \right]\,\,, \nonumber\\
\langle A;B \rangle_z &=& \int^{\infty}_0 dt\,\, \e^{izt}\, (A(t);B)\,\,,
\ea
with $A(t)=\exp(iHt/\hbar)\,A\,\exp(-iHt/\hbar)$ and $\dot A ={i \over
  \hbar} [H,A]$ , furthermore we used the abbreviation
\be
\label{14}
\hat{J}_{k} =  \epsilon^{-1}(k, \omega)\,\,J_k\,\,.
\ee
The correlation functions can be evaluated by standard many particle methods
such as perturbation theory for thermodynamic Green functions. In this
context the correlation functions containing $\hat{J}_{k}$ are
obtained from irreducible diagrams to Green functions containing
$J_{k}$, which do not disintegrate cutting only one interaction line.

The expression (\ref{11}) for the polarization function is very
general. Depending on the set of observables $\{B_n\}$, different
special cases are possible such as the Kubo formula or the Boltzmann
equation to be discussed in the following section. It is also possible
to include two-particle nonequilibrium correlations \cite{R,RR} if an appropriate
set of $B_n$ is chosen. We will work out here an approach to the
dielectric function which is closely related to the Chapman-Enskog
approach to the electrical conductivity.

\section{Moment expansion of the polarization function}

Up to now, $B_n$ was not specified. It is an advantage of the approach
given here that different levels of approximations can be constructed,
depending on the use of different sets of $B_n$. If no
finite order perturbation expansion of the correlation functions is
performed in evaluating the polarization function (\ref{11}), all
these different approaches are exact and should give identical
results. However, evaluating the correlation functions within
perturbation theory, different results for the polarization function
are expected using different sets of $B_n$. As has been shown for the
electrical conductivity \cite{R,RR}, results from finite
order perturbation theory are the better the more relevant observables
are considered.

A simple example for a relevant observable $B_n$ characterizing the
nonequilibrium state of the system is the current density (\ref{10}),
\be
\label{15}
B_{n}=J_{k}\,\,.
\ee
During this paper, we will treat this approach in detail. The current
density is related to the lowest moment of the distribution function.
Possible extensions to more general sets of relevant observables are
discussed at the end of this section.

In the approach given by Eq. (\ref{15}), we have
\be
\label{16}
\Pi(k, \omega) =-\,{i k^2 \beta \Omega_0 \over \omega}\,\, \frac {(J_{k};J_{k})
\,\,(J_{k};\hat J_{k})}{M_{JJ}}\,\,,
\ee
with
\be
\label{17}
M_{JJ}=-i \omega\,\, (J_{k};J_{k})+\langle \dot J_{k}; \dot
J_{k} \rangle_{\omega+i \eta} -
\frac{\langle \dot J_{k}; J_{k} \rangle_{\omega+i \eta}}
{\langle J_{k};J_{k} \rangle_{\omega+i \eta}}
\,\,\langle J_{k};\dot J_{k }\rangle_{\omega+i \eta}\,\,.
\ee
For the derivation we used the property
\be
\label{18}
(\dot A; B) = {i \over \hbar \beta}\,\, {\rm Tr} \{[A,B^+] \rho_0\}
\ee
(for proving perform the integral in the definition (\ref{13})) so that
$(J_k;\dot J_k)= {i \over \hbar \beta}\, {\rm Tr} \{ [J_k, J_{-k}]
\rho_0 \}=0$.

Applying integration by part (\ref{57}), the expression (\ref{16}) can
be rewritten as
\be
\label{19}
\Pi(k, \omega) =-\,\,{i k^2 \beta \Omega_0 \over \omega}\,\,
{(J_{k};\hat J_{k})\,\, \langle J_{k};J_{k} \rangle_{\omega +i \eta} \over
(J_{k};J_{k})-  \eta\,\, \langle  J_{k};
J_{k} \rangle_{\omega +i \eta} }
\ee
Performing the limit $\eta \rightarrow 0$  , for finite values of the
correlation function $ \langle  J_{k};
J_{k}\rangle_{\omega +i \eta} $
we obtain the simple result
\be
\label{20}
\Pi(k, \omega) = -\,\,{i k^2 \beta \Omega_0 \over \omega}\,\,
\langle J_{k};\hat J_{k} \rangle_{\omega +i \eta}
\ee
which is also denoted as the Kubo formula for the polarization
function. Similarly, the Kubo formula can also be obtained from more
general sets of observables $\{B_n\}$. A direct derivation of the
Kubo formula is obtained from Appendix A, Eq. (\ref{62}), if the set
of relevant observables $B_n$ is empty. Different approaches based on
different sets of relevant observables $B_n$ are formally equivalent
as long as no approximations in evaluating the correlation functions
are performed.

However, expressions (\ref{16} ) and (\ref{20} ) are differently
suited to perform perturbation expansions. For this we consider the
static conductivity $\sigma=\sigma(0,0)$ which follows from
\be
\label{21}
\sigma(k,\omega) = i\,\, {\omega \over k^2}\,\, \Pi(k,\omega)
\ee
in the limit $k \rightarrow 0 ,\,\, \omega \rightarrow 0$   .

Comparing the Kubo formula
\be
\label{22}
\sigma = \beta \Omega_0\,\,
\langle J_{0};\hat J_{0} \rangle_{i \eta}
\ee
with the result according to (\ref{16}),
\be
\label{23}
\sigma = \beta \Omega_0\,\,
 {(J_{0};J_{0})\,\, (J_{0};\hat J_{0}) \over \langle \dot J_{0}; \dot
J_{0} \rangle_{i \eta} -
\langle \dot J_{0}; J_{0} \rangle_{i \eta}\,\,
{\langle J_{0};J_{0} \rangle}^{-1}_{i \eta}\,\,
\langle J_{0};\dot J_{0}\rangle_{i \eta}}\,\,,
\ee
it is evident that perturbation theory cannot be applied to (\ref{22})
because in zeroth order this expression is already diverging. In
contrast, (\ref{23}) allows for a perturbative
expansion. For instance, in Born approximation the Faber -- Ziman result
for the electrical conductivity is obtained. The expression $
\sigma^{-1} \sim \langle \dot J_{0}; \dot J_{0} \rangle_{i \eta}$ is
also known as the force--force correlation function expression for the
resistivity. More precisely, the resistivity should be  given in terms
of stochastic forces which are related to the second term in the
denominator of Eq. (\ref{23}), see also Eq. (\ref{58}) in App. A. The
applicability of correlation functions for the inverse transport
coefficients has been widely discussed \cite{Z}.

The approach to the dielectric function given in the
present paper is based on the choice (\ref{15}) for the set of
relevant observables and may be considered as the generalization of
the force--force correlation function method for the electrical
resistivity to the dielectric function.
Possible extensions of the set of relevant observables have been
investigated in evaluating the dc conductivity in Ref. \cite{R}
and will be considered in evaluating the dielectric function in a
forthcoming paper \cite{RW}. 

\section{Evaluation of correlation functions}

Within the generalized linear response approach, the polarization
function is given in terms of correlation functions which, in general,
are elements of matrices. Within a quantum statistical approach, the
correlation functions are related to Green functions which can be
evaluated by diagram techniques. This has been discussed in detail in
the case of the static electrical conductivity \cite{R} and will not be
detailed here. Instead, we will consider only the lowest orders of
perturbation theory (Born approximation).

In the case considered here, the relevant observable $J_k$ (\ref{10}) is
given by a single particle observable. The correlation functions
occuring in (\ref{16}) will contain the operators
$n^c_{p,k} = c^+_{p-k/2} c^{}_{p+k/2}$  and $\dot n^c_{p,k} = -(i\hbar
  p_z k/m_c)\, n^c_{p,k} + v^c_{p,k}$, with
\be
\label{26}
 v^c_{p,k} = {i \over \hbar} \sum_{d,p',q} V_{cd}(q) \left[
 c^+_{p-k/2-q}\,\, d^+_{p'+q}\,\, d^{}_{p'}\,\, c^{}_{p+k/2} -
 c^+_{p-k/2}\,\, d^+_{p'+q}\,\, d^{}_{p'}\,\, c^{}_{p+k/2+q} \right]\,\,.
\ee

To evaluate the correlation functions, we perform a perturbation
expansion with respect to the interaction $V$, see App. B. In addition
to the zeroth order terms, which reproduce the RPA result, we consider
the Born approximation. Up to second order with respect to $V$ we have
\be
\label{16a}
\Pi(k, \omega) =-{i \beta \Omega_0 {k^2 \over \omega} \langle J_{k};J_{k}
\rangle^{(0)}_{\omega+i \eta} \over 1 +
\sum_{cd,pp'} {\hbar^2 \over \Omega_0^2}\,\, {e_c e_d \over m_c m_d}
\,\,p_zp'_z { \langle v^c_{p,k};
v^d_{p',k} \rangle^{(0)}_{\omega+i\eta} \over (J_{k};J_{k})^{(0)}}
\left[ {1\over \eta - i \omega
  + i {\hbar \over m_d} p'_z k} + {1\over \eta - i \omega
  + i {\hbar \over m_c} p_z k} - { \langle
  J_{k};J_{k} \rangle^{(0)}_{\omega+i\eta} \over (J_{k};J_{k})^{(0)}}
\right] } \,\,.
\ee
The evaluation of the correlation functions for the non-degenerate
case is shown in App. B. We obtain the following expression
\be
\label{16b}
\Pi (k,\omega) = - { \beta \sum_c e^2_c\, n_c \,[1+z_cD(z_c)] \over 1
  - i {\omega \over k^2} {e_e^2 e_i^2 \over (4 \pi \epsilon_0)^2}\, n_e n_i
{\mu_{ei}^{1/2}  \over ( k_BT)^{5/2}} {2 (2 \pi)^{1/2} \over
 \sum_c e^2_c\, n_c \,/m_c} \int_0^\infty dp \,\, \e^{-p^2} \left( \ln {\lambda -1
\over \lambda +1} + {2 \over \lambda+1} \right)
W(p) }
\ee
with
\ba
W(p)&=& {2 \over
    3} \,\,p\,\,\left(  {e_e \over m_e}-{e_i \over m_i}  \right)^2
{\sum_c e^2_c\, n_c \,[1+z_cD(z_c)] \over \sum_c e^2_c\, n_c \,/m_c}\nonumber\\
&-& {M^{1/2}_{ei} \over  \mu^{1/2}_{ei}} \left(  {e_e \over m_e}-{e_i \over
  m_i}  \right) \int_{-1}^1 dc\,\, c \left[ e_e D(z_{ei}-\sqrt{{m_i \over
    m_e}} c p) + e_i
D(z_{ei}+\sqrt{{m_e \over m_i}} c p) \right]\,\,.
\ea
Here, $z_{ei} = {\omega \over k} \sqrt{{M_{ei} \over 2 k_BT}} $,
$z_c = {\omega \over k} \sqrt{{m_c \over 2 k_BT}} $,
$\lambda(p) = (\hbar^2 \kappa^2)/(4 \mu_{ei} k_BT p^2) +1$ ,
and
\be
\label{31a}
D(z)={1 \over \sqrt{\pi}} \int_{-\infty}^\infty \e^{-x^2} {dx \over
  x-z-i\eta} =i \sqrt{\pi} \e^{-z^2} \left[ 1+{\rm Erf}(i z)
  \right]\,\,
\ee
denotes the Dawson integral. Note that a statically screened potential
was used in (\ref{26}) to obtain a convergent collision integral, the
screening parameter is given by $\kappa^2 = \sum_c e_c^2 n_c/(\epsilon_0
k_BT)$. From (\ref{16b}) it is immediately seen that the RPA result is
obtained in the limit of vanishing interactions, $W(p)=0$.


\section{Results for hydrogen plasmas}

The expression (\ref{16b}) for the polarization function is simplified
for a system consisting of protons and electrons, where $e_i=-e_e,
\,\,n_i=n_e$, and $m_i/m_e=1836$:
\ba
\label{16c}
&&\epsilon (k,\omega) = 1+ { e^2\, n  \over\epsilon_0 k_BT k^2}
[2+z_eD(z_e)+z_iD(z_i)]
\left[ 1-i {\omega \over k^2} {e^4 \over (4 \pi \epsilon_0)^2} n
  {\mu^{1/2} \over (k_BT)^{5/2}} 2 (2 \pi)^{1/2} \int_0^\infty dp \,\,
  \e^{-p^2} \left( \ln {\lambda -1 \over \lambda +1} + {2 \over
    \lambda+1} \right)\right. \nonumber\\
&& \left. \times \left\{ {2 \over 3} \,\,p\,\, [2+z_eD(z_e)+z_iD(z_i)]- \left(
{M_{ei} \over \mu_{ei}} \right)^{1/2}
 \int_{-1}^1 dc\,\, c \left( D(z_{ei}-\sqrt{{m_i \over
m_e}} c p) - D(z_{ei}+\sqrt{{m_e \over m_i}} c p) \right) \right\} \right]^{-1}
\ea

We first discuss the limiting case of small $k$. For $k \ll
\omega \sqrt{m_e/(2k_BT)}$ we use the expansion
\be
D(z)=i \sqrt{\pi} \e^{-z^2}-{1 \over z}- {1 \over 2 z^3} \pm \dots
\ee
so that after expanding also with respect to
$cp/z_{ei}$ we have
\be
\label{16d}
\epsilon (0, \omega) = 1-{\omega_{\rm pl}^2 \over \omega^2 +i
\omega/\tau}
\ee
with $\omega_{\rm pl}^2=e^2n/(\epsilon_0 \mu_{ei})$ and
\be
\tau=
%
{(4 \pi \epsilon_0)^2 \over e^4}
{(k_BT)^{3/2} \mu_{ei}^{1/2} \over n} {3 \over 4 (2 \pi)^{1/2} } \left[
 \int_0^\infty dp\,\, p\,\, \e^{-p^2} (\ln {\lambda-1 \over
  \lambda+1} +{2 \over \lambda+1}) \right]^{-1}
\ee
According to (\ref{7}), the dc conductivity
\be
\sigma(0,\omega \rightarrow 0)=\omega_{\rm pl}^2 \epsilon_0 \tau
\ee
is obtained, what coincides with the Faber-Ziman formula at finite
temperatures \cite{R}.

On the other hand, in the limiting case of small $\omega$ we use
for $\omega \ll \sqrt{2 k_BT/m_i} k$ the expansion
\be
D(z)=i \sqrt{\pi} \e^{-z^2} - 2 z +{4 \over 3} z^3 \pm \dots
\ee
and obtain
\be
\label{D}
\lim_{k\rightarrow 0} \lim_{\omega \rightarrow 0} \epsilon (k, \omega)
= 1+ {\kappa^2 d \over -i \omega + d k^2} \left( 1 + i {\omega
    \over 2 k} \sqrt{{ \pi m_i \over 2 k_BT}} \right)
\ee
with
\be
d^{-1} = - {e^4 \over (4 \pi \epsilon_0)^2} n
{4 (2 \pi)^{1/2} \mu_{ei}^{1/2} \over(k_BT)^{5/2}}
 \int_0^\infty {dp  \over p} \,\,\e^{-p^2} (\ln {\lambda-1 \over
  \lambda+1} +{2 \over \lambda+1})
\ee
Here, in evaluating the last expression of (\ref{16c}), also
$z_{ei}+\sqrt{m_e/m_i} cp$ is considered as a small quantity,
whereas $z_{ei}-\sqrt{m_i/m_e} cp$ is large in the region of
relevant $p$.
For small values  $k < \sqrt{2 k_BT/(\pi m_i)} 2/d$, the second
term in the numerator of
(\ref{D}) can be neglected, and the diffusion type form of
$\epsilon(k, \omega)$ is obtained, see \cite{Kli}.

As an example, a dense plasma is considered with parameter values $T=
50$ eV and $n_e=3.2\,\,10^{23}$ cm$^{-3}$. Such parameter values have
been reported recently in laser produced high-density plasmas by Sauerbrey et al.,
see \cite{S}. We will use Rydberg units so that $T=3.68$ in Ryd and
$n_e=0.0474$ in $a_B^{-3}$. At these parameter values, the plasma
frequency is obtained as $\omega_{\rm pl}= 1.54$, and the
screening parameter as $\kappa =0.805$.

First we discuss the dependence of the dielectric function on
frequency for different values of $k$, see Figs. 1-4. For large values
of $k$ our result for the dielectric function coincides with the RPA
result. At decreasing $k$ strong deviations are observed. Both
the RPA expression as well as the
expression (\ref{16c}) for the dielectric function
fulfill important relations such as the Kramers-Kronig relation
and the condition of total screening. The validity
of the sum rule
\be
\int_0^\infty \omega\, {\rm Im} \epsilon (k, \omega)\,d \omega =
{\pi \over 2} \omega_{\rm pl}^2
\ee
is checked by numerical integration. The RPA result
coincides with the exact value $\omega_{\rm pl}^2 \pi /2=3.74$ to be
compared with expression (\ref{16c}) which gives 3.74 at $k=1$, 3.75 at
$k=0.1$, 3.71 at $k=0.01$ and 3.74 at $k=0.001$. The small
deviations are
possibly due to numerical accuracy.

To investigate the behavior at small $k$, we give a log-log plot of
Im$\epsilon(k, \omega)$ as function of $\omega$ for different values $k$ in
Fig. 5. For $\omega > \sqrt{2 k_BT/m_e} k =3.84 k$ the Drude-like behaviour
(\ref{16d}) is clearly seen, with $\tau =8.36$.

Considering the limit of small $\omega$, a log-log plot of
Im$\epsilon(k, \omega)$ as function of $k$ for different values $\omega$ is shown
in Fig. 6. The diffusion behavior (\ref{D}) occurs for $k < \sqrt{2 k_BT/(\pi
m_i)} =0.00732$ at $k > \sqrt{m_i/(2 k_BT)} \omega = 11.17 \omega$
with $d=13.8\,$. Altogether the numerical evaluation of the general expression
(\ref{16c}) for the dielectric function confirms the validity of the simple limiting
formulae (\ref{16d}) and (\ref{D}).

In this paper we have focussed the discussion only to the properties of $\epsilon(k,
\omega)$. Related quantities such as $\epsilon^{-1}(k,\omega)$ will be
investigated in a forthcoming paper \cite{RW}. The parameter values for density
and temperature can be extended to other nondegenerate plasmas like
ordinary laboratory plasmas or the solar plasma. This has been done with
results showing the same qualitative behavior of the expression
(\ref{16c}) in comparison with the RPA expression, but at
shifted values of $k$ and $\omega$.

\section{Conclusions}

An expression for the dielectric function of Coulomb systems is
derived which is consistent with the Chapman-Enskog approach to
the dc
conductivity. For a two-component plasma, explicit calculations have
been performed in the lowest moment approach. In Born
approximation, expressions are given which allow the determination of
$\epsilon(k, \omega)$ in an analytical way. It is shown that 
general relations such as sum rules are fulfilled as well as the dc
conductivity is obtained in the form of the Ziman-Faber result.

We performed exploratory calculations to illustrate how the
generalized linear response approach works. Obviously an improvement
of the results can be obtained if i) the Born approximation is
improved including higher order of perturbation theory, ii) higher
moments of the single-particle distribution are taken into account.
Both points have been discussed for the limiting case of the dc
conductivity \cite{R}, where a virial expansion of the inverse
conductivity was given.

A four moment approach will be presented in a subsequent paper
\cite{RW} where also the comparison with the Kubo approach and
computer simulations are discussed. Within the approach given here it
is also possible to treat the degenerate case. Work in this direction
is in progress.

\section*{Acknowledgement}
The author is indebted to August Wierling for many helpful discussions
and to Arne Schnell for help in performing the computer
calculations.

\section*{Appendix A: Generalized linear response theory}

To construct the nonequilibrium statistical operator $\rho(t)$ we use
the density matrix approach \cite{Z,RC}. Characterizing the
nonequilibrium state of the system by the mean values $\langle A_n
\rangle^t$ of a set of relevant observables $\{A_n\}$, the generalized
Gibbs state
\be
\label{38}
\rho_{\rm rel}(t) = \e^{ -S(t)/k_B}\,\,,
\ee
where
\be
\label{39}
{1 \over k_B} S(t) = \Phi(t)+\sum_n \alpha_n(t)\,A_n
\ee
is the entropy operator and
\be
\label{40}
\Phi(t)= \ln {\rm
  Tr}\exp \left\{-\sum_n \alpha_n(t)A_n \right\}
\ee
is the Massieu-Planck function, follows from the maximum of the entropy
\be
\label{41}
\langle S(t) \rangle^t =-k_B\, {\rm Tr}\{\rho_{\rm rel}(t)\,\ln
\rho_{\rm rel}(t)  \}
\ee
at given mean values
\be
\label{42}
{\rm Tr}\{ A_n \,\,\rho_{\rm rel}(t) \} = \langle A_n\rangle^t\,\,.
\ee
The thermodynamic parameters (Lagrange multipliers) $\alpha_n(t)$ are
determined by the self-consistency conditions (\ref{42}) and will be
evaluated within linear response theory below.

The relevant statistical operator (\ref{38}) does not solve the von
Neumann equation, but it can serve to formulate the correct boundary
conditions to obtain the retarded solution of the von Neumann
equation. Using Abel's theorem, the nonequilibrium statistical
operator \cite{Z} is found with the help of the time evolution operator
$U(t,t')$,
\be
\label{43}
i \hbar\, {\partial \over \partial t} U(t,t') =H_{\rm tot}(t)\,\, U(t,t');
\qquad U(t',t')=1,
\ee
as
\be
\label{44}
\rho(t)=\eta \int_{-\infty}^t dt'\,\, \e^{-\eta(t-t')}\,\, U(t,t')\,\,
\rho_{\rm  rel}(t')\,\, U(t',t)\,\,,
\ee
where the limit $\eta \rightarrow 0$ has to be taken after the
thermodynamic limit. Partial integration of (\ref{44}) gives
\be
\label{45}
\rho(t)=\rho_{\rm rel}(t)+\rho_{\rm irrel}(t)
\ee
with
\be
\label{46}
\rho_{\rm irrel}(t)=-\int_{-\infty}^t dt'\,\, \e^{-\eta(t-t')}
U(t,t')\,\,\left\{ {i \over \hbar} \left[H_{\rm tot}(t')\,,\, \rho_{\rm
  rel}(t')\right] +{\partial \over \partial t'}\, \rho_{\rm
  rel}(t')\right\}  U(t',t).
\ee
The self-consistency conditions (\ref{42}) which determine the
Lagrange multipliers take the form
\be
\label{47}
{\rm Tr}\{ A_n\,\, \rho_{\rm irrel}(t) \} = 0\,\,.
\ee

For a weak external field $U_{\rm ext}$, the system remains near
thermal equilibrium described by $\rho_0$ (\ref{5}), so that
$\rho(t)$ (\ref{45}) can be expanded up to the first order with respect
to  $U_{\rm ext}$. For this we specify the set of relevant observables
$\{A_n\}$ as $\{H, N_c, B_n(\vec r) \}$ (note that summation over $n$
in (\ref{39}) also means integration over $\vec r$) and the
corresponding Lagrange
parameters $\{\alpha_n \}$ as $\{ \beta, -\beta \mu_c, - \beta
\phi_n(\vec r,t) \}$,
\be
\label{48}
{1 \over k_B} S(t) = \Phi(t) + \beta H - \beta \sum_c \mu_c\, N_c -
\beta \sum_n \int d^3 r\,\, \phi_n(\vec r,t)\,\, B_n(\vec r)\,\,.
\ee
Expanding the nonequilibrium statistical operator up to first order
with respect to $U_{\rm ext}$ and $\phi_n(\vec r,t)$ it is convenient
to use the Fourier representation\footnote{In general we have
  $\phi_n(\vec r,t) =  \sum_{k'} \int {d \omega' \over 2 \pi}\,
  \e^{i(\vec k' \vec r -
    \omega' t)} \phi_n(\vec k', \omega')$ and $B_{n,k'} = \int d^3 r
\, B_n(\vec r)\, \e^{-i \vec k' \vec r}$. The selfconsistency equations
(\ref{47}) must be fulfilled for any time $t$ so that $\omega'=
\omega$ follows.  Furthermore, the
equilibrium correlation functions ${\rm Tr} ( A_k B_{k'}^+\rho_0)$ do
not vanish only if $k'=k$ so that $\langle A_k(\eta-i\omega);B_{k'}
\rangle \sim \delta_{kk'}$. The well-known property of linear response
that only such fluctuations are induced where the
wave vector and frequency coincide with the external potential is a
consequence of homogeneity in space and time.} so that
\be
\label{49}
\int d^3r\,\, \phi_n(\vec r,t)\,\, B_n(\vec r) = \phi_n(\vec k,
\omega)\,\, \e^{-i \omega t}\, B_n^+ + {\rm c.c.}
\ee
with
\be
\label{50}
\phi_n(\vec r,t)= \e^{i(\vec k \vec r-\omega t)}\,\,\phi_n(\vec k,\omega)
\,\,,\qquad B_{n}= \int  d^3 r \,\, B_n(\vec r)\,\, \e^{-i \vec k \vec r}\,\,.
\ee
The contributions to (\ref{45}) are
\be
\label{51}
\rho_{\rm rel}(t)= \rho_0 +  \e^{-i \omega t} \int_0^\beta
d\tau \sum_n  B_n^+(i \hbar \tau)\,\,\phi_n(\vec k,\omega)\,\,\rho_0 + {\rm
  c.c.}
\ee
%
and, applying the Kubo identity
\be
\label{52}
[A,\rho_0]=\int_0^\beta d\tau\,\,\e^{-\tau H}\,\, [H,A]\,\, e^{\tau
  H}\,\,\rho_0\,\,,
\ee
we find
\ba
\label{53}
&&\rho_{\rm irrel}(t)=- \int_{-\infty}^t dt' \,\,\e^{-\eta(t-t')}\,\,
\e^{-i \omega t'} \int_0^\beta d\tau \left\{
\sum_{c,p} e_c\,\, \dot n^c_{p,-k}(t'-t+i\hbar \tau)\,\, U_{\rm
  ext}(\vec k,\omega)
\right. \nonumber\\
&&\left. + \sum_n
\left[ \dot B_{n}^+(t'-t+i \hbar \tau)-i \omega B_{n}^+(t'-t+i \hbar
  \tau) \right] \, \phi_n(\vec k,\omega) \right\} \rho_0  + {\rm c.c.}
\ea
Inserting this result in the self-consistency conditions (\ref{47})
we get the response equations
\be
\label{54}
-\,\, \langle B_m;A \rangle_{\omega+i \eta}\,\, U_{\rm eff}(\vec k,\omega) =
\langle B_m;C \rangle_{\omega+i \eta}
\ee
with the correlation functions defined by (\ref{13}),
\be
\label{55}
A=\sum_{c,p}e_c\,\, \dot n^c_{p,k} = i k \Omega_0\, \hat{J}_{k}\,\,,
\ee
and
\be
\label{56}
C= \sum_n
\left[ \dot B_{n}-i \omega B_{n} \right] \phi_n(\vec k,\omega)\,\,.
\ee

To make the relation between the response equations (\ref{54}) and
the Boltzmann equation more closely, see \cite{R}, we introduce the
'stochastic' part of forces applying partial integrations
\be
\label{57}
 \langle A;B \rangle_z={i \over  z} \left[(A;B) +
 \langle \dot A;B \rangle_z \right]={i \over z} \left[(A;B) -
 \langle  A;\dot B \rangle_z \right]
\ee
so that (\ref{54}) can be rewritten as
\ba
\label{58}
&&-i k \Omega_0 \left( B_m; \hat J_{k} \right) U_{\rm eff}(k,\omega)
= \frac{ \left( B_m; J_{k} \right) + \langle \dot B_{m};
 J_{k} \rangle_{\omega+i \eta} -\langle \dot B_{m};
 J_{k} \rangle_{\omega+i \eta} }{ \langle B_{m};
 J_{k}\rangle_{\omega+i \eta}}\,\,
\langle B_m;C \rangle_{\omega+i \eta} \nonumber \\
&&=\left(B_{m};C \right)+\Big\langle \left[ \dot B_m -
\frac{\langle \dot B_m;J_k \rangle_{\omega+i \eta}}{\langle  B_m
 ;J_k \rangle_{\omega+i \eta}}\,\, B_m \right];\left[
C -  \frac{\langle  B_m ;C
  \rangle_{\omega+i \eta}} {\langle  B_m
  ;J_k \rangle_{\omega+i \eta}}\,\,J_{k} \right]\Big\rangle_{\omega+i \eta}
\ea
Then, we find the following form for the response
equations
\be
\label{59}
-\, i k \Omega_0 \,\,M_{m0}\,\, U_{\rm eff}(k,\omega) = \sum_n
M_{mn}\,\, \phi_n(k,\omega)
\ee
with
\be
\label{60}
M_{m0}= \left( B_m ;\hat J_k \right)
\ee
and
\ba
\label{61}
&& M_{mn} = \left( B_m;[\dot B_n-i \omega B_n] \right)
+ \Big\langle \left[ \dot B_m - \, \frac{\langle \dot B_m ;J_k
  \rangle_{\omega+i \eta}}{\langle B_m ; J_k  \rangle_{\omega+i \eta}}\,\,
 B_m \right];[ \dot B_n-i \omega B_n]
\Big\rangle_{\omega+i \eta}\,\,.
\ea

The system of equations (\ref{59}) can be solved applying Cramers
rule. Then, the response parameters are represented as a ratio of two
determinants.

With the solutions $\phi_n$ the explicit form of $\rho (t)$ is known,
and we can evaluate mean values of arbitrary observables. In
particular, we are interested in the evaluation of $\langle J_k
\rangle^t \exp (i \omega t)$ to calculate the polarization function
(\ref{9}) using (\ref{51}), (\ref{53}),
\ba
\label{62}
\langle J_k \rangle^t \,\,\e^{i \omega t} &=&\beta \sum_n \left\{ (J_k;B_{n}) -
 \langle J_{k};[\dot B_{n}-i \omega B_n] \rangle_{\omega+i \eta}
\right\} \phi_n (\vec k,\omega) \nonumber \\
&-& i k \Omega_0 \beta\,\, \langle J_{k};
\hat{J}_k \rangle_{\omega+i \eta}\,\, U_{\rm eff}(\vec k, \omega)\,\,.
\ea
If $J_k$ can be represented by a linear combination of the
relevant observables $\{B_n\}$, we can directly use the
selfconsistency conditions (\ref{42}) and have
\be
\label{63}
\langle J_k \rangle^t\,\, \e^{i \omega t} = {\rm Tr} \left[ J_k\,\, \rho_{\rm rel}
(t) \right]\,\, \e^{i \omega t}\,.
\ee
Comparing with (\ref{62}) we see that the remaining terms on the rhs
of (\ref{62}) compensate due to the response equations (\ref{59}).
After expanding $\rho_{\rm rel}(t)$ up to first order in $\phi_n(\vec k,\omega)$,
Eq. (\ref{51}), we have
\be
\label{64}
\langle J_k \rangle^t \,\,\e^{i \omega t} = \beta \sum_n
( J_{k};B_n) \,\,\phi_n(\vec k,\omega)\,\,.
\ee
Inserting the solutions for $\phi_n$ in the form of determinants, we
get the same result as obtained if we expand the numerator
determinant (\ref{11}) with respect to its first row.

\section*{Appendix B: Evaluation of the collision term in Born
approximation}


Let us first consider the lowest order of perturbation theory where we
have for the correlation functions
\ba
\label{27}
&&(n^d_{p,k};n^c_{p',k}) = \hat f^c_{p,k}\,\, \delta_{pp'}\,\, \delta_{cd}
\nonumber\\
&&\langle n^d_{p,k};n^c_{p',k} \rangle_{\omega+i\eta} = (\eta -i
\omega+ i \hbar p_z k/m_c )^{-1} \hat f^c_{p,k}\,\, \delta_{pp'}\,\,
\delta_{cd}\,\,,
\ea
where
\be
\label{28}
\hat f^c_{p,k}=(\beta \hbar^2 p_z k/m_c)^{-1}(f^c_{p-k/2}-f^c_{p+k/2})
\ee
Notice that $\lim_{k \rightarrow 0} f^c_{p,k} =f^c_p = \{ \exp [ \beta
(E^c_{p}-\mu_c)] +1 \}^{-1}$.
In the classical limit where the Fermi function can be replaced by the
Maxwell distribution. we have in lowest order with respect to the
Coulomb interaction
\be
\label{jj}
(J_{k};J_{k})^{(0)} = {k_BT \over \Omega_0} \sum_c {e_c^2 \over m_c}\,\, n_c\,\,,
\ee
\be
\label{30}
\langle J_{k};J_{k} \rangle^{(0)}_{\omega+i\eta} = - i\,\, {\omega \over k^2}\,\,
{1 \over \Omega_0} \sum_c e^2_c\,\, n_c\,\, [1+z_cD(z_c)]
\ee
with $z_c = {\omega \over k} \sqrt{{m_c\over 2 k_B T}}$ and
\be
\label{31}
D(z)={1 \over \sqrt{\pi}} \int_{-\infty}^\infty \e^{-x^2} {dx \over
  x-z-i\eta}\,\,.
\ee
Furthermore we have
\ba
\label{32}
\langle \dot J_{k};J_{k} \rangle^{(0)}_{\omega+i\eta}& =&
-\,\, {k_BT \over \Omega_0} \sum_c {e_c^2 \over m_c}\, n_c
- {\omega^2 \over k^2}\,\,
{1 \over \Omega_0} \sum_c e^2_c\, n_c\, [1+z_cD(z_c)]\nonumber\\
&= & -\,\,\langle J_{k};\dot J_{k} \rangle_{\omega+i\eta}\,\,,
\ea
\be
\label{33}
\langle \dot J_{k};\dot J_{k} \rangle^{(0)}_{\omega+i\eta} =
-i \omega\, {k_BT \over \Omega_0} \sum_c {e_c^2 \over m_c}\, n_c
-i\, {\omega^3 \over k^2}\,
{1 \over \Omega_0} \sum_c e^2_c\, n_c\, [1+z_cD(z_c)]\,\,,
\ee
so that from Eq. (\ref{16}) the random phase approximation (RPA)
\be
\label{34}
\Pi^{(0)} (k,\omega) = -  \beta \sum_c e^2_c\, n_c \,[1+z_cD(z_c)]
\ee
is obtained.

After we have considered the collisionless plasma, we will now treat
the general case of an interacting system where the correlation
functions have to be evaluated with the full Hamiltonian (\ref{1}). The
evaluation of equilibrium correlation functions for an interacting
many-fermion system can be performed within perturbation theory such
as a Green function approach, and many-particle effects can be treated
in a systematic way. We will give here the lowest order contribution
with respect to the screened Coulomb interaction (Born approximation),
a systematic treatment of higher orders can be done as indicated in
\cite{R} for the case of static conductivity.

In the numerator of (\ref{16}), the higher order expansion for 
$(J_{k};J_{k})$ lead
to the replacement of the occupation numbers $f^c_p$ for the free
fermion gas by the occupation numbers in an interacting fermion
gas. This corrections in Born approximation can be given as shift of
the single-particle energies and can be replaced by a shift of the
chemical potential.

We will investigate here the collision terms where the Born
approximation leads to essential contributions. For this we use the
relations (proof by partial integration (\ref{57}))
\be
\label{35}
\langle n^c_{p,k};v^d_{p',k} \rangle_{\omega+i \eta} = (\eta -i
\omega+ i \hbar p_z k/m_c )^{-1}\left[ ( n^c_{p,k};v^d_{p',k} ) +
\langle v^c_{p,k}; v^d_{p',k} \rangle_{\omega+i \eta}\right]\,\,,
\ee
\be
\label{36}
\langle v^c_{p,k};n^d_{p',k} \rangle_{\omega+i \eta} = (\eta -i
\omega+ i \hbar p'_z k/m_d )^{-1}\left[ ( v^c_{p,k};n^d_{p',k}) -
\langle v^c_{p,k}; v^d_{p',k} \rangle_{\omega+i \eta}\right]\,\,,
\ee
and find considering only the interaction in the collision terms
\ba
\label{37}
M_{JJ}&& =
{(J_{k};J_{k})^2 \over \langle J_{k};J_{k} \rangle_{\omega+i\eta}}
+ \sum_{cd,pp'} {\hbar^2 \over \Omega_0^2}\,\, {e_c e_d \over m_c m_d}
\,\,p_zp'_z\,\, \langle v^c_{p,k};
v^d_{p',k} \rangle_{\omega+i\eta} \nonumber\\
&&\times \left\{-1+{(J_{k};J_{k}) \over \langle
  J_{k};J_{k} \rangle_{\omega+i\eta}} \left[ {1\over \eta - i \omega
  + i \hbar p'_z k/m_d} + {1\over \eta - i \omega
  + i \hbar p_z k/m_c} \right] \right\}\,\,.
\ea

In the Born approximation for the frequency and wave vector dependent
collision term we take the evolution operator due to the
noninteracting part $H^0$ of the Hamiltonian (\ref{1}) so that the
correlation functions are immediately evaluated using Wick's
theorem. Dropping single-particle exchange terms what can be justified
for the Coulomb interaction in the low-density limit, we find
\ba
\label{vv}
&&\langle v^c_{p,k}(\eta- i \omega); v^d_{p',k} \rangle = - {\pi
  \over \hbar}\sum_{c'p''q} {\exp(\beta \hbar \omega)-1 \over \beta
  \hbar \omega} V_{cc'}(q) f^{c'}_{p''+q} (1-f^{c'}_{p''})\nonumber\\
&&\times \left\{ f^c_{p+k/2-q}(1-f^c_{p-k/2}) \delta (E^c_{p+k/2-q} +
E^{c'}_{p''+q} - E^c_{p-k/2} - E^{c'}_{p''} - \hbar \omega)\right. \nonumber\\
&&\times \left[
V_{cc'}(-q) \delta_{cd}( \delta_{p',p-q} - \delta_{p',p}) +
V_{c'c}(-k+q) \delta_{c'd}( \delta_{p',p''-k/2+q} -
\delta_{p',p''+q/2}) \right] \nonumber\\
&&-  f^c_{p+k/2}(1-f^c_{p-k/2+q}) \delta (E^c_{p+k/2} +
E^{c'}_{p''+q} - E^c_{p-k/2+q} - E^{c'}_{p''} - \hbar \omega)
\nonumber\\
&& \left. \times \left[
V_{cc'}(-q) \delta_{cd}( \delta_{p',p} - \delta_{p',p+q}) +
V_{c'c}(-k+q) \delta_{c'd}( \delta_{p',p''-k/2+q} -
\delta_{p',p''+q/2}) \right] \right\}\,\,.
\ea

We evaluate the matrix element $M_{JJ}$, Eq. (\ref{37}) in Born
approximation to obtain the polarization function $\Pi (k, \omega)$,
Eq. (\ref{16}).
Using (\ref{17}), (\ref{jj}), (\ref{30}), (\ref{37})
we introduce
\be
\label{65}
R= {(J_{k};J_{k})^{(0)} \over \langle J_{k}; J_{k}
\rangle^{(0)}_{\omega+i\eta}}= i k_BT\, {k^2 \over \omega}\,\,
{\sum_c e_c^2 \,n_c/m_c \over
\sum_c e_c^2\, n_c\, [1+z_c D(z_c)]}
\ee
and find the perturbation expansion $M_{JJ}= M_{JJ}^{(0)}+M_{JJ}^{(1)}$,
where
\be
\label{m0}
M_{JJ}^{(0)}
=R\,\,(J_{k};J_{k})^{(0)}\,\,,
\ee
\be
\label{67}
M_{JJ}^{(1)}= {\hbar^2 \over \Omega_0^2} \sum_{cd,pl} {e_ce_d \over m_cm_d}
\,\,p_z l_z \,\,\langle v^c_{p,k}; v^d_{l,k} \rangle_{\omega+i\eta}
\left\{ -1 + R \left[ {1 \over \eta- i \omega+ i \hbar p_z k / m_c}
+ {1 \over \eta- i \omega+ i \hbar l_z k / m_d} \right] \right\}
\ee
Evaluating the correlation functions $
\langle v^c_{p,k}; v^d_{l,k} \rangle_{\omega+i\eta}$ in Born
approximation (\ref{vv}), we have for small $k$, $\omega$
\ba
\label{68}
M_{JJ}^{(1)} & = & 2\, {\pi \hbar \over \Omega_0^2} \sum_{lpq} V^2_{ei}(q)
f^e_p f^i_l\,\, \delta (E^e_{p+q}+E^i_{l-q}-E^e_p-E^i_l)\,\, q_z
\left( {e_e \over m_e}-{e_i \over m_i} \right)
\nonumber\\ & \times & \left\{ \left( {e_e \over m_e} p_z +
{e_i \over m_i} l_z \right) - 2 R
\left( {p_z \over i \hbar k p_z/m_e - i \omega + \eta}\,\, {e_e \over
m_e} + {l_z \over i \hbar k l_z/m_i - i \omega + \eta}\,\,
{e_i \over m_i} \right) \right\}\,\,.
\ea

The further evaluation is done with introducing total
and relative momenta $\vec P=\vec p+\vec l, \,\, \vec p'=(m_i
\vec p-m_e \vec l)/M_{ei},\,\, \vec p''=\vec p' + \vec q$,
$M_{ei}=m_e+m_i, \,\,\mu_{ei}^{-1}=m_e^{-1}+m_i^{-1}$ so
that
\ba
\label{69}
M_{JJ}^{(1)} & = & 2\,\, {\hbar \pi \over \Omega_0}\,\,
{e^2_e e^2_i \over \epsilon_0^2}\,\,
n_e \left({2 \pi \hbar^2 \over m_e k_BT} \right)^{3/2}
n_i \left({2 \pi \hbar^2 \over m_i k_BT} \right)^{3/2}
{1 \over (2 \pi)^9}\,\, {2 \mu_{ei} \over \hbar^2} \int d^3P \int
d^3p' \int d^3 p'' \nonumber\\
& \times & \e^{-{\hbar^2 P^2 \over 2
M_{ei} k_BT}} \e^{-{\hbar^2 p'^2 \over 2 \mu_{ei} k_BT}} \delta
(p'^2-p''^2) {1 \over ((\vec p'-\vec p'')^2+\kappa^2)^2}
(p''_z-p'_z) \left( {e_e \over m_e}-{e_i \over
m_i} \right) \nonumber \\
& \times & \left\{ p'_z \left( {e_e \over m_e}-{e_i \over
m_i} \right) - 2 R {M_{ei} \omega \over i \hbar^2 k^2}
\left( {e_e  \over P_z+ {M_{ei}
\over m_e} p'_z - {M_{ei} \omega \over \hbar k} - i \eta} +
{e_i \over P_z - {M_{ei}
\over m_i} p'_z - {M_{ei} \omega \over \hbar k} - i \eta}
\right) \right\}\,\,.
\ea

Furthermore we introduce dimensionless variables $\hbar P (2
M_{ei} k_BT)^{1/2},\,\,$ $\hbar p' (2 \mu_{ei} k_BT)^{1/2},\,\,$
$\lambda = (\hbar^2 \kappa^2)/(4 \mu_{ei} k_BT p'^2) +1$ and
spherical coordinates $p'=\{ p' (1-c^2)^{1/2},0,p' c\},\,\,$ $
p''=\{ p'' (1-z^2)^{1/2} \cos \phi, p'' (1-z^2)^{1/2} \sin \phi, p''z
\}$ and perform the integral over $\phi$ according to
\be
\label{70}
\int_0^{2 \pi} d \phi {1 \over [\lambda-cz- \sqrt{1-c^2}
\sqrt{1-z^2} \cos \phi]^2} = 2 \pi {\lambda -cz \over
(\lambda^2-1+c^2-2 \lambda c z + z^2)^{3/2}}
\ee
so that
\ba
\label{71}
M_{JJ}^{(1)}& = &
{1 \over \Omega_0} n_e n_i {e_e^2 e_i^2 \over \epsilon_0^2}
\left( {\mu_{ei} \over (2 \pi)^3 k_BT} \right)^{1/2} {1 \over 2}
\int_0^\infty {1 \over p'} dp' \e^{-p'^2} \int_{-1}^1 dc
\int_{-1}^1 dz {1 \over \pi^{3/2}}\int d^3P \e^{-P^2} \nonumber\\
& \times &  {\lambda -cz \over
(\lambda^2-1+c^2-2 \lambda c z +z^2)^{3/2}} (z-c) \left\{ p'^2 c \left( {e_e \over m_e}-{e_i \over
m_i} \right)^2 \right. \nonumber\\
& + &  i R p'  \left( {e_e \over m_e}-{e_i \over
m_i} \right)  { \omega \over k_BT k^2} \sqrt{{M_{ei} \over
\mu_{ei}}} \left.
\left[ {e_e \over P_z + \sqrt{{m_i \over m_e}} p' c- {\omega \over k}
\sqrt{{M_{ei} \over 2 k_BT}}} +
{e_i \over P_z - \sqrt{{m_e \over m_i}} p' c- {\omega \over k}
\sqrt{{M_{ei} \over 2 k_BT}}} \right] \right\}\,\,.
\ea
Now, the integrals over $z$ and $P$ can be performed. Using
\be
\label{72}
\int_{-1}^1 dz \,\, {\lambda -cz \over
(\lambda^2-1+c^2-2 \lambda c z +z^2)^{3/2}}\,\, (z-c) =
c \left( \ln {\lambda -1 \over \lambda +1} + {2 \over \lambda
+1} \right)\,\,,
\ee
we finally find
\ba
\label{73}
&&M_{JJ}^{(1)} = {1 \over \Omega_0}\, n_e n_i\,\, {e_e^2 e_i^2 \over
\epsilon_0^2} \left( {\mu_{ei} \over 2 k_BT} \right)^{1/2} {1
\over 4 \pi^{3/2}} \int_0^\infty dp\,\, \e^{-p^2} \left( \ln {\lambda -1
\over \lambda +1} + {2 \over \lambda+1} \right) \nonumber\\
&&\times \left\{{2 \over 3}\,\, p \left( {e_e \over m_e}-{e_i \over
m_i} \right)^2 + i R \left( {e_e \over m_e}-{e_i \over
m_i} \right) { \omega \over k_BT k^2} \sqrt{{M_{ei} \over \mu_{ei}}}
\int_{-1}^1 dc\,\,c \left[ e_e D(z_{ei}-\sqrt{{m_i \over m_e}} c p) + e_i
D(z_{ei}+\sqrt{{m_e \over m_i}} c p) \right] \right\}
\ea
with $z_{ei} = {\omega \over k} \sqrt{{M_{ei} \over 2 k_BT}} $.
Together with (\ref{m0}), (\ref{jj}), this result can be inserted in expression
(\ref{16}) to evaluate $\Pi(k,\omega)$.

Figure captions:

Fig.1: $\epsilon(k, \omega)$ as function of $\omega$ (in Ryd/$\hbar$) at 
$k=1/a_B$
for a hydrogen plasma, $n_e = 3.2 \,\,10^{23}$cm$^{-3}, \,\,\,T=50$ eV.\\
a: Re $\epsilon$, b: Im $\epsilon$.\\
broken line: RPA, full line: first moment Born approximation.

Fig.2: The same as Fig.1 for $k=0.1/a_B$.

Fig.3: The same as Fig.1 for $k=0.01/a_B$.

Fig.4: The same as Fig.1 for $k=0.001/a_B$.

Fig.5: Im $\epsilon(k, \omega)$ as function of $\omega$ for
different $k$.

Fig.6: Im $\epsilon(k, \omega)$ as function of $k$ for
$\omega = 0.000001$ Ryd/$\hbar$.

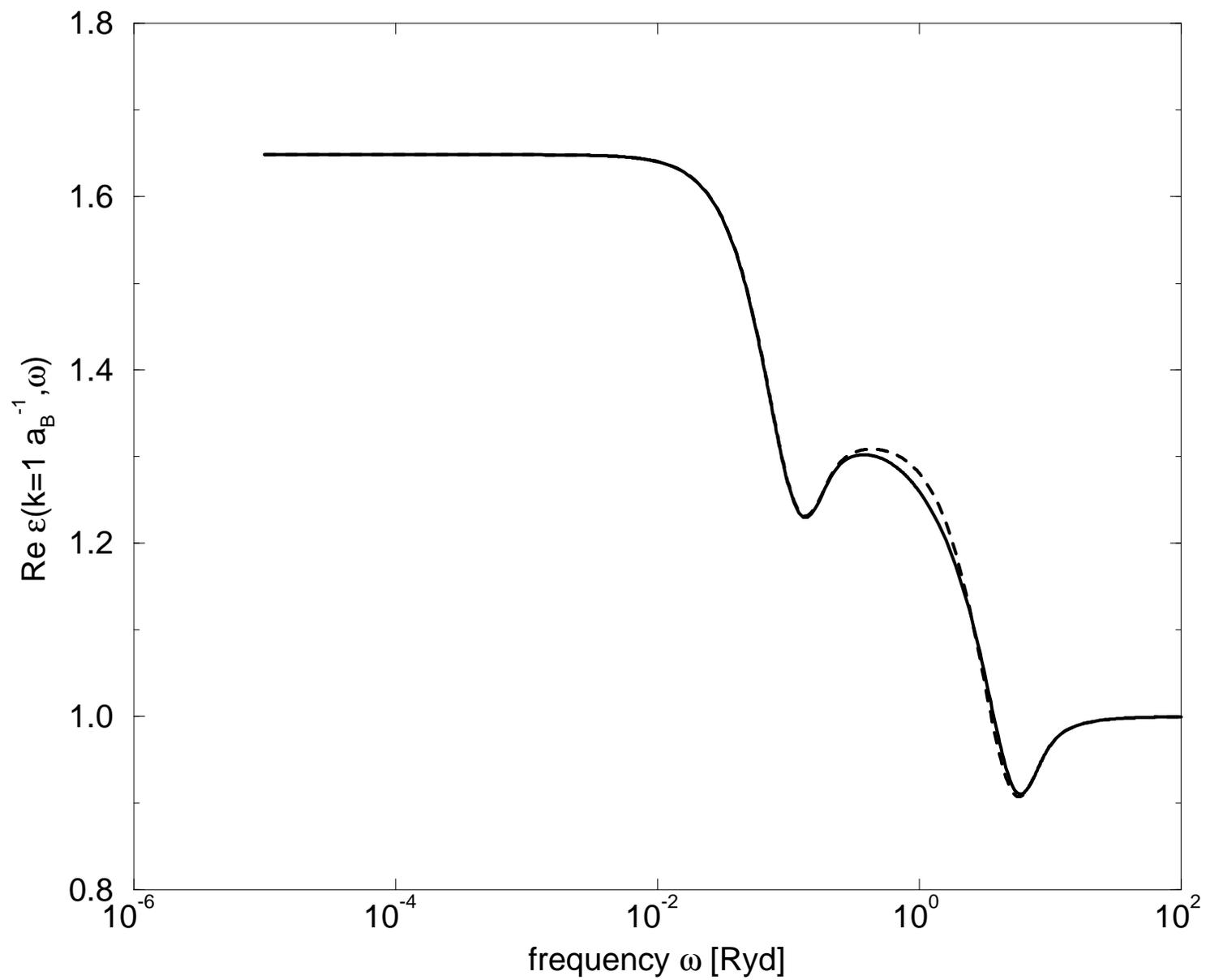
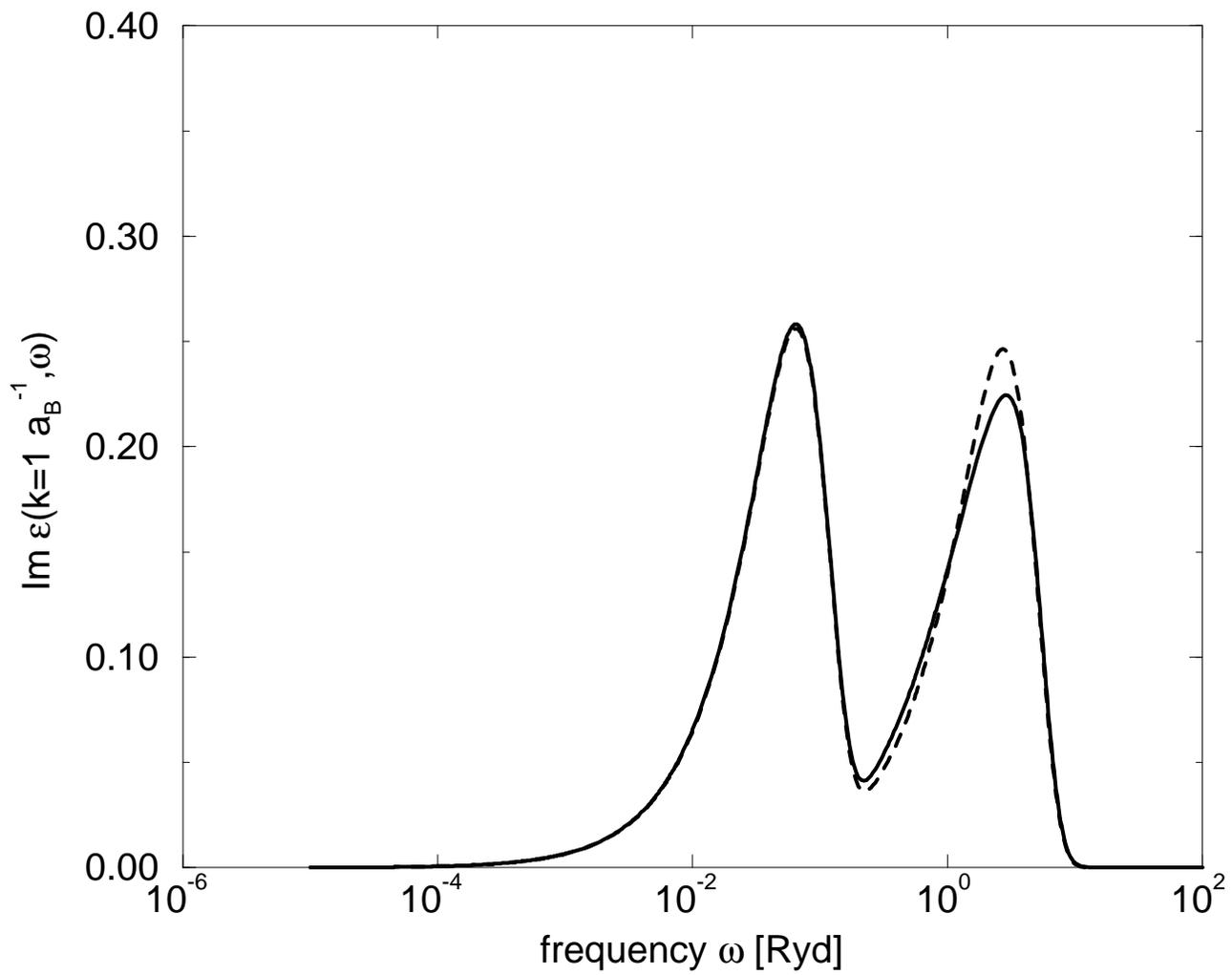
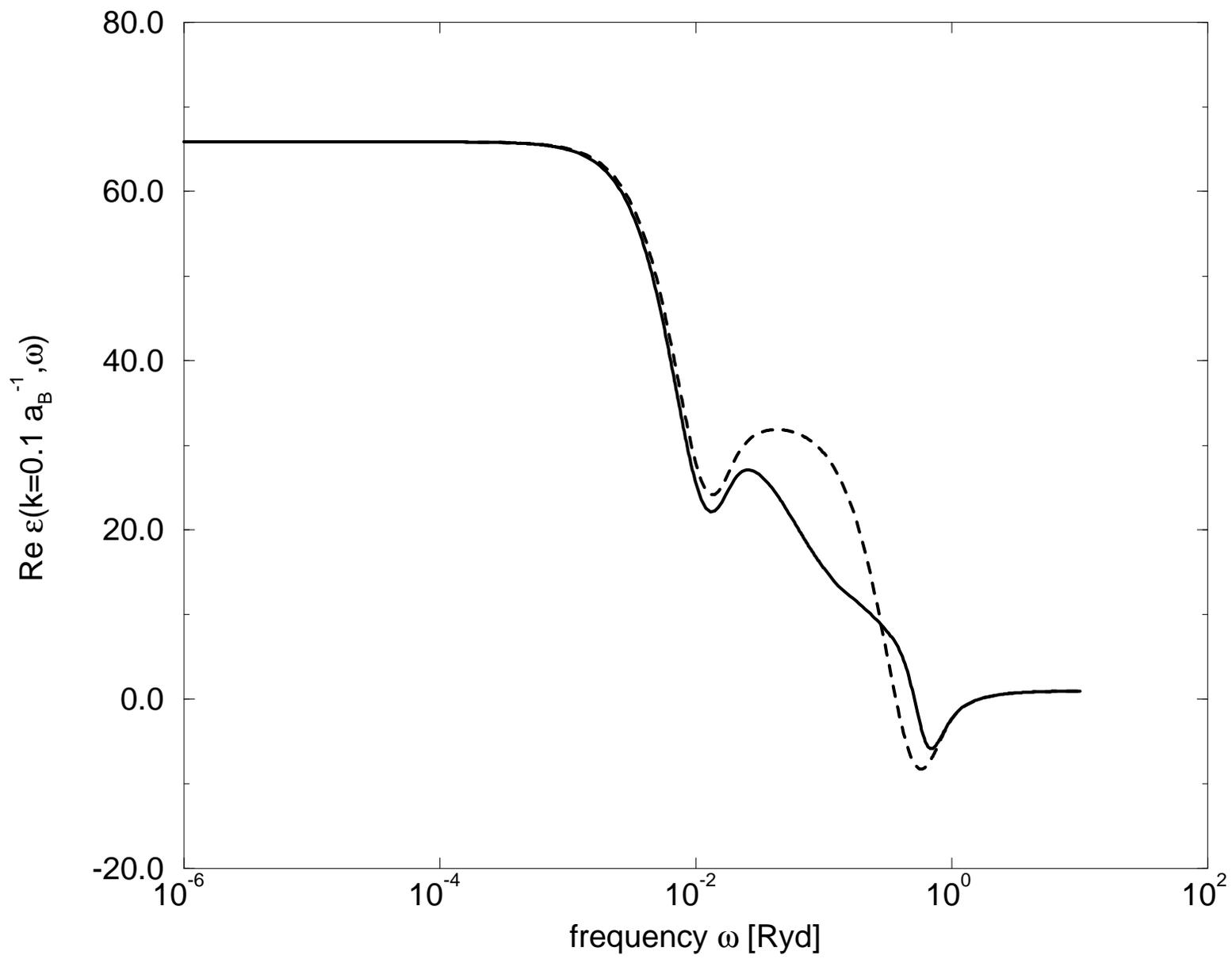
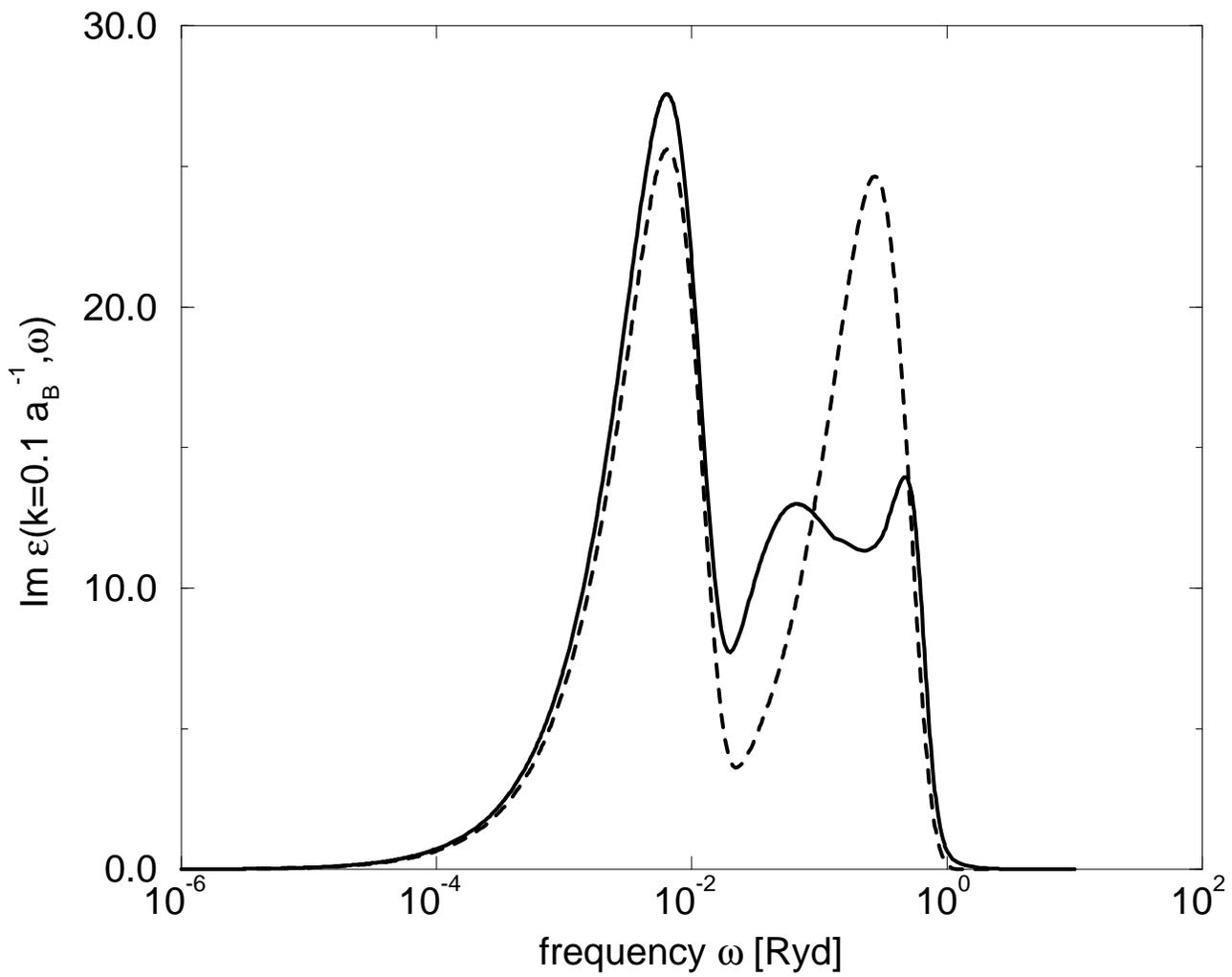
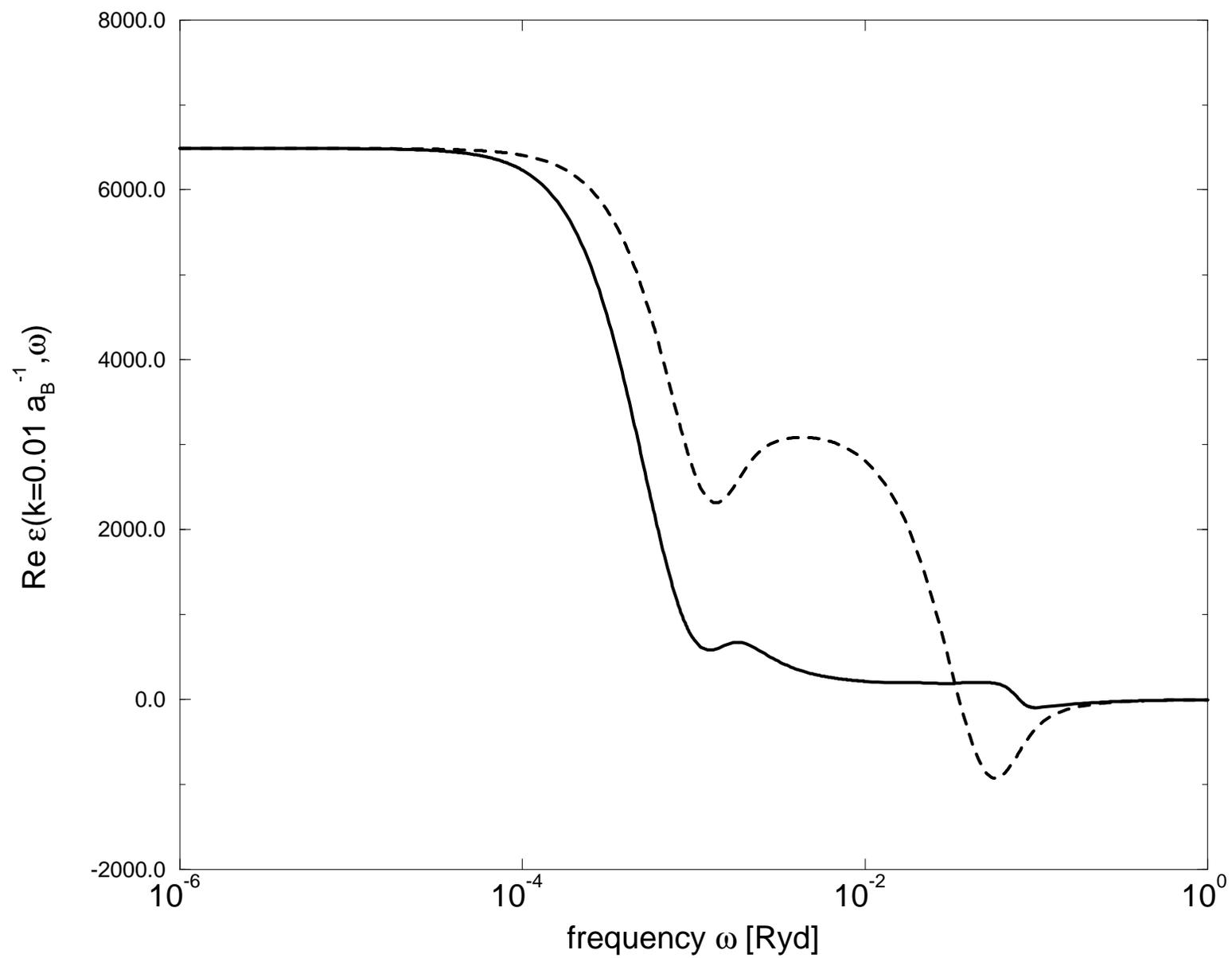
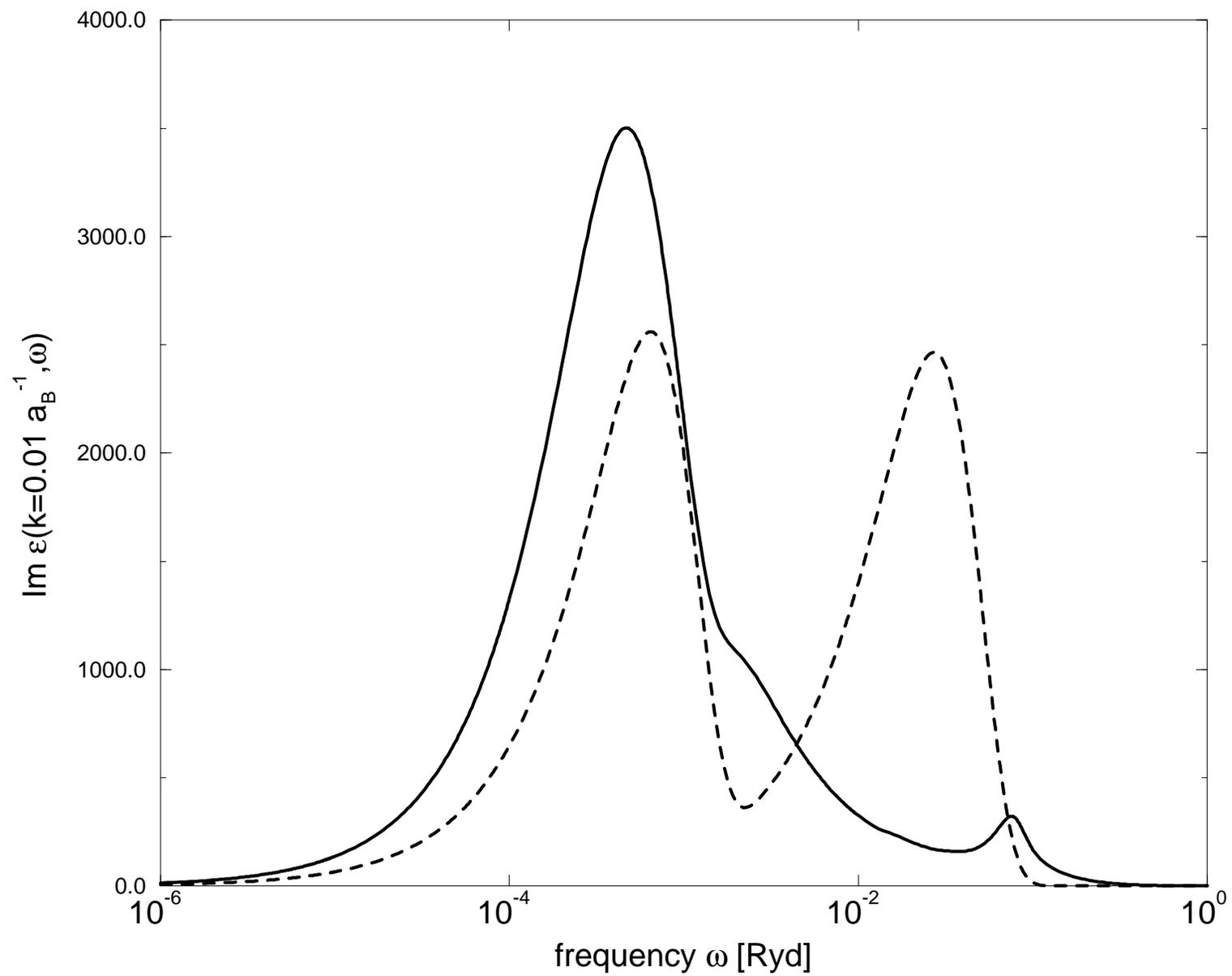
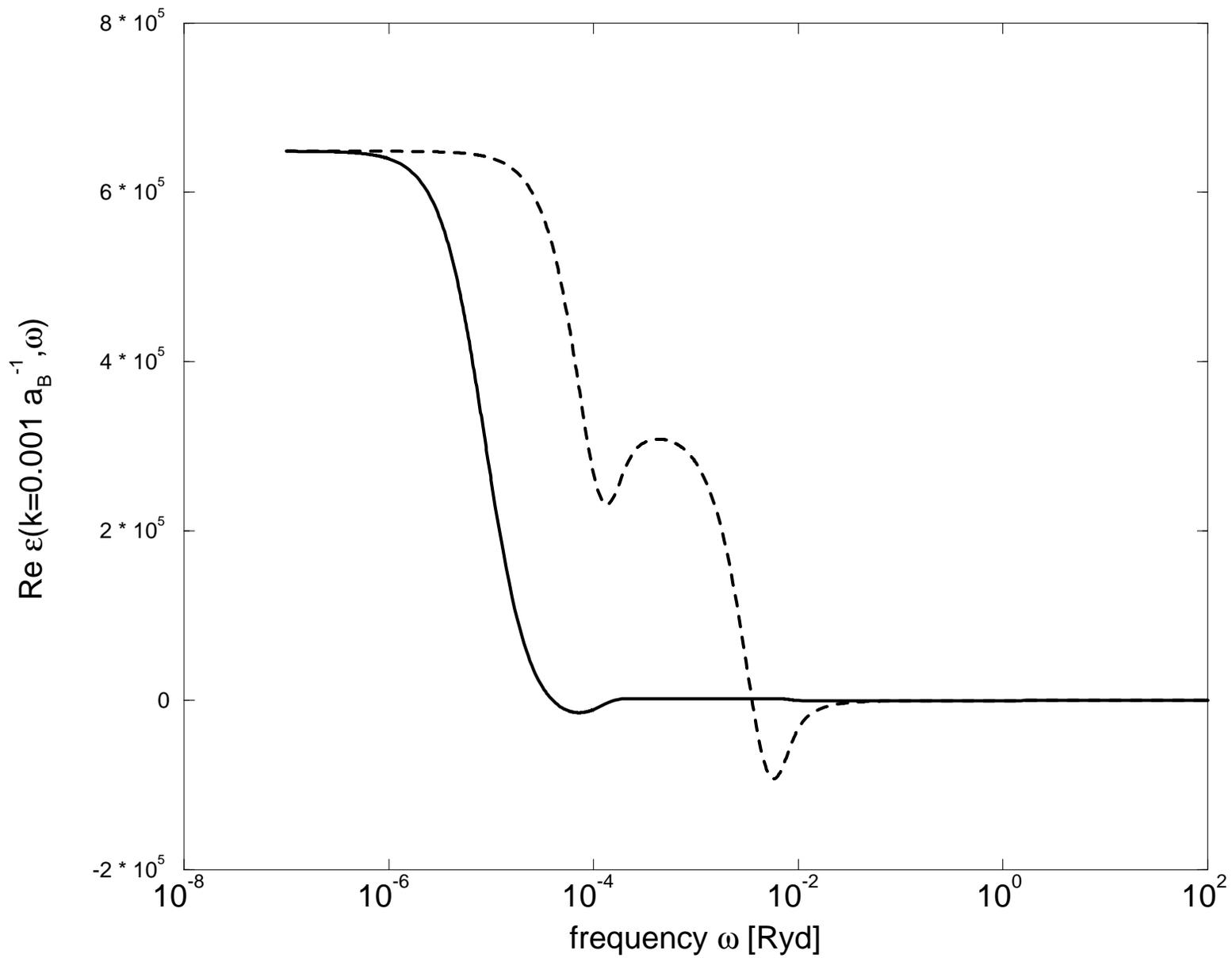
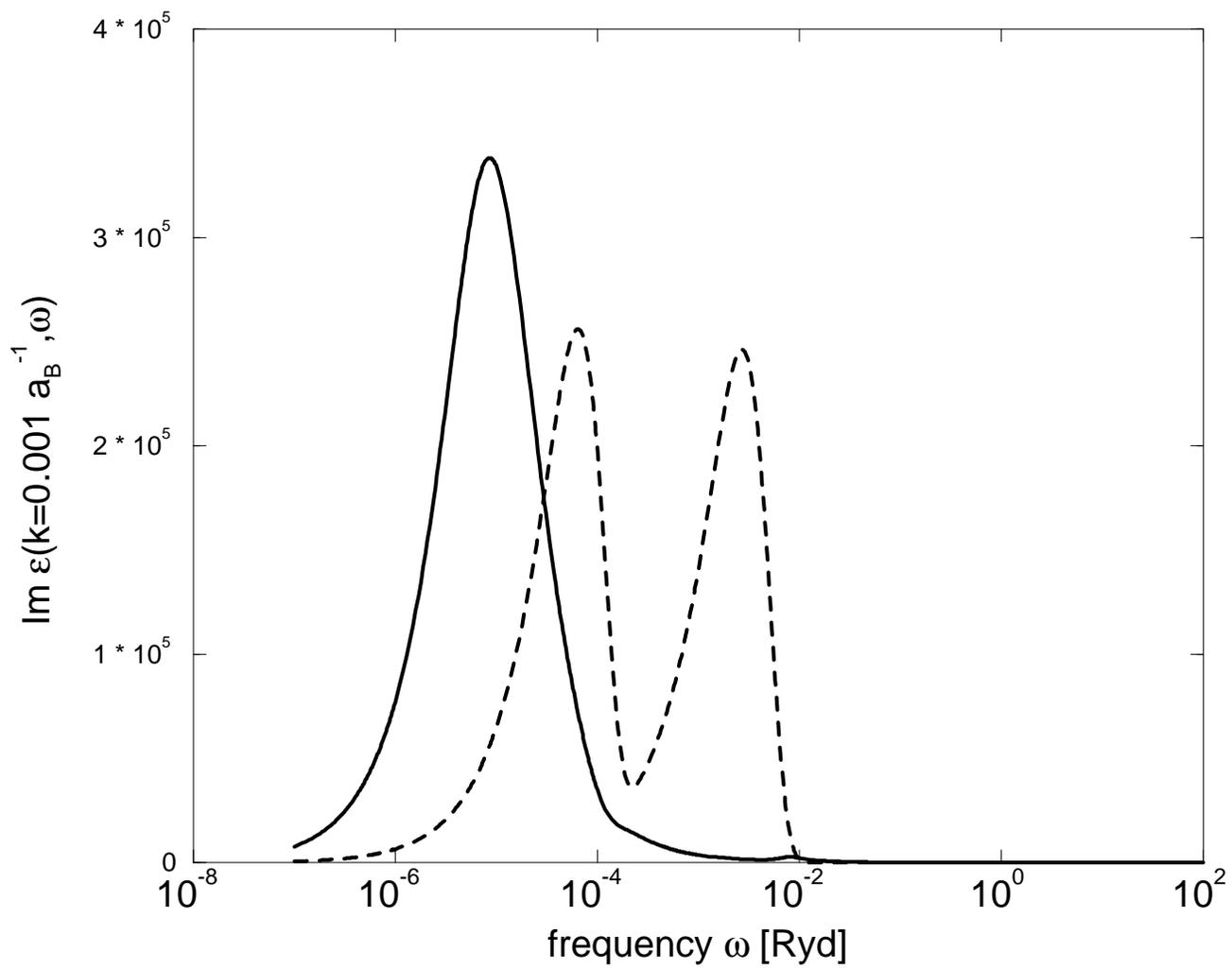
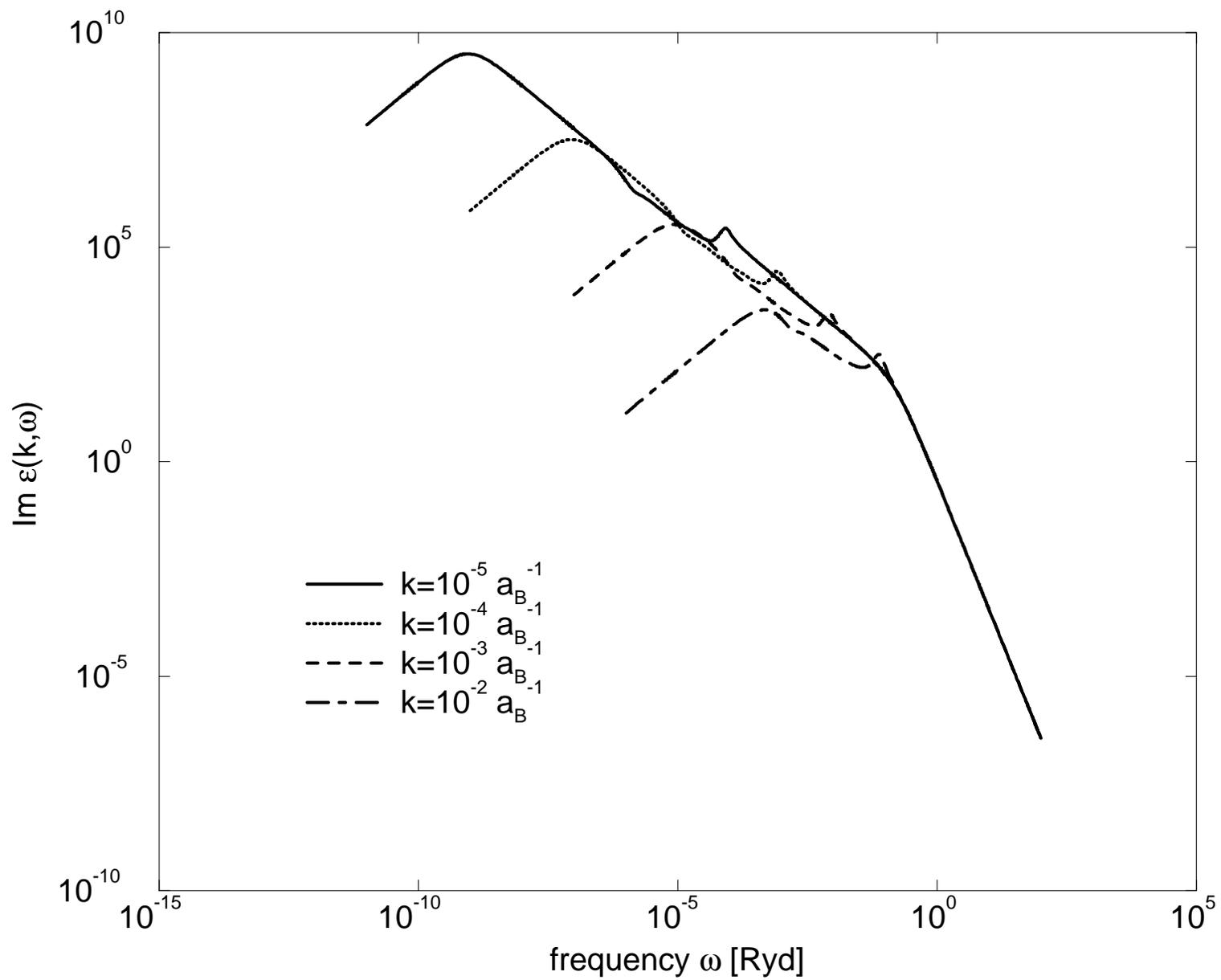
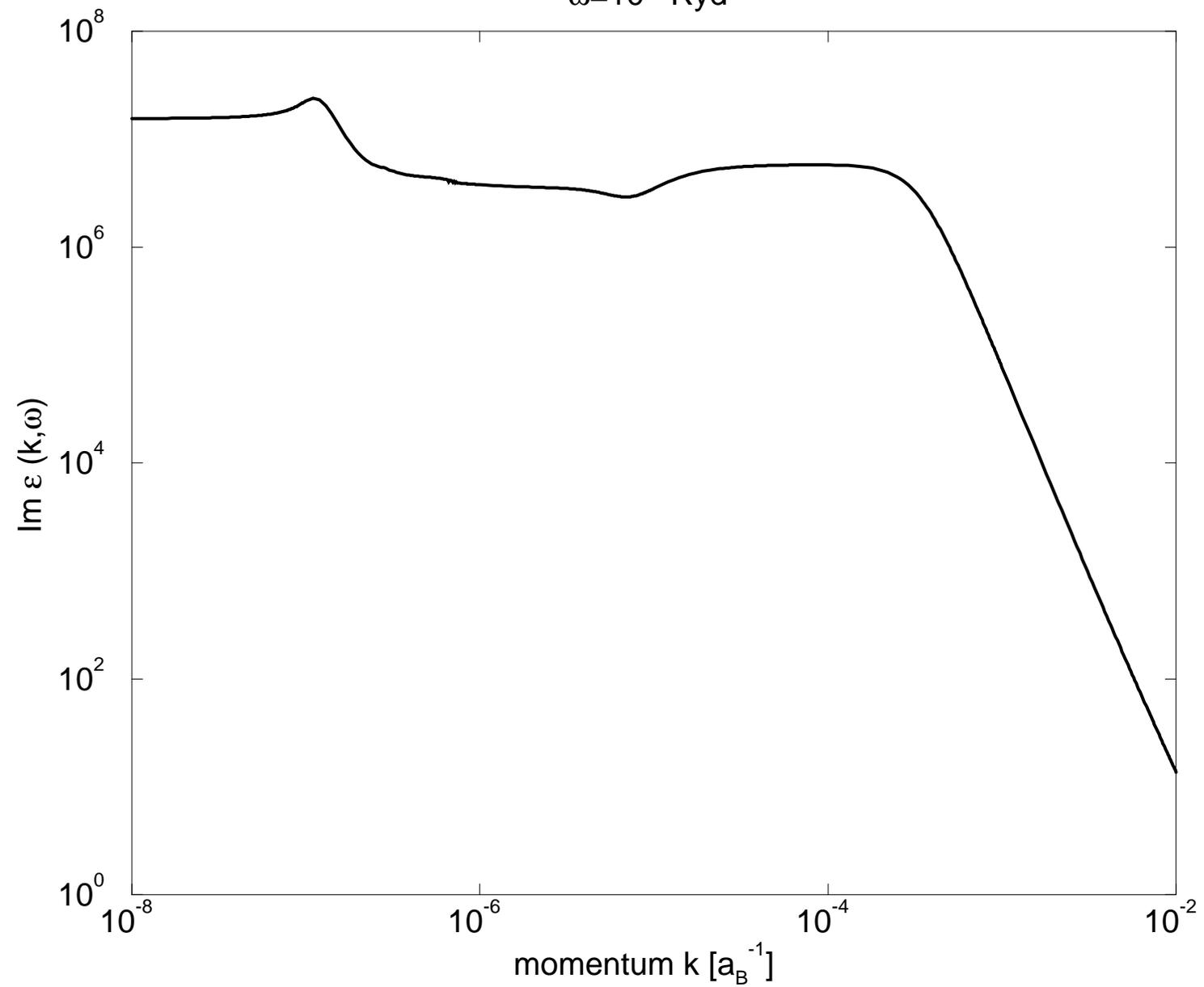
\end{document}